# The role of IT ambidexterity, digital dynamic capability and knowledge processes as enablers of patient agility: an empirical study

Rogier van de Wetering[1*] & Johan Versendaal[1]

[1]Department of Information Sciences, Open University of the Netherlands, 6419 AT Heerlen, The Netherlands

*Correspondence: rogier.vandewetering@ou.nl

**Abstract**

There is a limited understanding of IT's role as a crucial enabler of patient agility and the department's ability to respond to patient's needs and wishes adequately. This study's objective is to contribute to the insights of the validity of the hypothesized relationship between IT resources, practices and capabilities, and hospital departments' knowledge processes and the department's ability to adequately sense and respond to patient needs and wishes, i.e., patient agility. This study conveniently sampled data from 107 clinical hospital departments in the Netherlands and uses structural equation modeling for model assessment. IT ambidexterity positively enhances the development of a digital dynamic capability. Likewise, IT ambidexterity also positively impacts the hospital department's knowledge processes. Both digital dynamic capability and knowledge processes positively influence patient agility. IT ambidexterity promotes taking advantage of IT resources and experiments to reshape patient services and enhance patient agility.

# Introduction

In the age of digital transformation, modern hospitals need to simplify their current care delivery processes and sustainable business models to contain the staggering rising healthcare costs and address the needs of the more engaged and informed patient. At the same time, hospitals need to adequately address the confluence of dynamic and unpredictable market forces in which they operate, optimally deploy and enable their IT assets, resources, and organizational, IT, and knowledge capabilities and focus on the state of the art patient service delivery [1-6]. Doctors and other medical professionals can use innovative information technology (IT) solutions and the available exponential volumes of patient-generated data—including the patient's medical history in a single, easy-to-find location—to enhance the quality of care delivery [7-9]. As a result, hospitals in this day and age need to deal with a myriad of substantial organizational, political, and technological challenges over the coming years, also in the process of fully leveraging digital technologies. [10, 11]. Emerging technologies like big data analytics, the Internet of Things, distributed ledger technologies, social media, AI, and cloud-based solutions are, in essence, more than promising. These innovative technologies can truly disrupt the quality of processes and services, the effectiveness of medical outcomes, the productivity of employees, and ultimately change lives [12-16]. Hospitals can now redefine their role in the hospital ecosystem so that the patient service quality and value might ultimately translate into substantial societal benefits [17].

Despite a wealth of attention for IT adoption and IT-enabled transformation in healthcare research [6, 18-25], there is still a limited understanding of the role of IT as a crucial enabler of organizational sensing and responding capabilities to address the needs, wishes, and requirement of patient adequately. [26-29]. Moreover, the extant scholarship has contended that IT could also hamper the process of gaining organizational benefits [30-33]. Understanding the facets that drive IT investments benefits is very valuable in clinical settings [34]. As can be gleaned from the above, substantial gaps remain in the extant literature. This paper, therefore, responds to two crucial limitations in the extant research. First, this current paper tries to unfold how hospital departments can develop the ability to simultaneously pursue 'exploration' and 'exploitation' in their management of IT practices, i.e., IT ambidexterity [35, 36], by practitioners often referred to as bimodal IT, see, e.g., [37, 38], to drive a hospitals' departments digital dynamic capability. This technical-oriented dynamic capability, in essence, represents the degree to which qualities and competencies are developed to manage innovative digital technologies for new exceptional and effective patient service development [39]. As such, this capability requires substantial undertakings toward embracing new digital technologies [39, 40]. Second, this study tries to unfold the complementary effect of IT ambidexterity and digital dynamic capability on hospital departments' knowledge processes and the departments' ability to adequately sense and respond to patient needs and wishes, i.e., patient agility. Healthcare processes require close collaboration between different clinical departments and disciplines, and IT is crucial in facilitating effective knowledge processes between key stakeholders (e.g., doctors, nurses) [8, 41-43]. Hence, IT-driven knowledge processes can enhance patient treatment processes and patient agility.

Gaining these insights is essential, as hospitals are currently very active in exploring their digital options, innovations and transforming their clinical processes and their interactions with patients using digital technologies [43, 44]. For instance, clinicians who use digital innovations in their clinical practice, e.g., mobile handheld devices and apps, can increase error prevention and improve patient-centered care [45-48]. In addition, digital options and innovations provide ways for clinicians to be more agile in their work, improve clinical communication, remotely monitor patients, and improve clinical decision-support [49-51] and hence improve the patient treatment process and quality of medical services [51, 52]. Moreover, recent scholarship advocates the deployment of knowledge assets, processes, and digital-driven sense and respond capabilities as a way of achieving higher-quality and patient-centered care and financial performance benefits in hospitals [46, 53, 54]. Moreover, Fadlalla and Wickramasinghe [55] argue that patient-centered[1] sensing, responding, and digital capabilities are crucial in facilitating high care quality.

These insights are also important for hospitals in the Netherlands, as Dutch hospitals are bound to care production agreements (i.e., so-called turnover ceilings) between hospitals and health insurers. The Dutch Healthcare Authority (NZa), an autonomous administrative authority falling under the Dutch Ministry of Health, Welfare, and Sport, oversees that these agreements focus more on patient quality and value creation than production. Therefore, more contract negotiations will be driven by focusing on the quality of care and patient value; achieving patient agility seems a valuable endeavor. Thus, this research tries to extend existing work on IT-enabled transformation in healthcare and does so by sufficiently capturing clinicians' attitudes toward IT ambidexterity, digital dynamic capability, knowledge processes, and patient agility of their respective hospital departments. In doing so, we adopt a practitioner-based approach [56, 57].

---

[1] Care that is respectful of and responsive to individual patient preferences, needs, and values.

Throughout this study, the dynamic capabilities framework is embraced [40, 58, 59]. As such, this study distinguishes between IT resources, a lower-order technical dynamic capability, and higher-order dynamic capabilities, i.e., knowledge processes and patient agility [40, 60-62].

To summarize, the study's main research questions are:

1. *How does IT ambidexterity lead to perceived patient agility of the hospital department?*

2. *What is the role of digital dynamic capability and knowledge processes in the process of converting the contributions of IT ambidexterity on the department's perceived patient agility enhancements?*

This study's IT-business value approach aligns with the industries' focus on operational and clinical excellence, patient-centered value, and a streamlined patient journey [63, 64].

Hence, this paper proceeds as follows. First, it reviews the theoretical development by highlighting key literature on IT resources and ambidexterity, the dynamic capabilities view, and organizational agility. Then, section 3 highlights the study's research model and associated hypotheses. Section 4 details the methods used in this study, after which section 5 outlines key results. This study ends with discussing the outcomes, including theoretical and practical contributions, and ends with concluding remarks.

# Theoretical background

## IT resources and IT ambidexterity

The ability to leverage IT resources is crucial to meet new business requirements adapt to changing market conditions, and obtain business benefits [3, 65-67]. In their overview of past and future research on digital transformation [68], identify the research theme on the contribution of organizational capabilities. Organizations need to pursue and make trade-offs between two seemingly opposing paths, i.e., the ability to adapt existing IT resources to the current business environment and demands and their focus on developing IT resources that contribute to long-term organizational benefits [36, 69]. The balance between these two objectives is referred to, in the literature, as ambidexterity [70-73]. The simultaneous engagement of 'exploration' and 'exploitation' by organizations will likely provide them with superior business benefits [70-73]. IT exploration concerns the organization's efforts to pursue new knowledge and IT resources [35, 69]. For instance, think about acquiring new IT resources (e.g., potential IT applications, critical IT skills) and an organization's ability to experiment with new IT management practices. On the other hand, IT exploitation is typically conceptualized as a construct that captures the degree to which organizations take advantage of existing IT resources and assets, e.g., the reuse of existing IT applications and services for new patient services and the reuse of existing IT skills. The extant literature has contended that IT ambidexterity can substantially impact process-based capabilities and organizational agility [35].

## Digital dynamic capability

Digital dynamic capabilities can be considered the "organization's skill, talent, and expertise to manage digital technologies for new product development" [39]. Hence, it can be conceived as an organization's ability to master digital technologies, drive digital transformations, and develop new innovative patient-centered services and products. Our study embraces a hierarchical capability view [35, 40, 58, 59]. Digital dynamic capability is conceptualized as a lower-order technical dynamic capability that organizations could embed and leverage in the process of developing higher-order dynamic organizational capabilities such as innovation ambidexterity, absorptive capacity, and organizational adaptiveness [40, 60, 61]. This current conceptualization is also in line with the previous scholarship. For instance, Benitez et al. (2018) conceptualized a flexible IT infrastructure as a dynamic capability that provides the organization with adequate responsiveness by enabling the business flexibility to sense and seize mergers and acquisition opportunities [74]. Likewise, Vitari et al. [75] argue that an organization's digital data genesis can be conceived as a dynamic capability. It allows identifying new digital data opportunities, generating and capturing data, and recombining existing data to adapt to changing market conditions. Also, Queiroz et al. [76] claimed the organization's ability to renew its IT application portfolio, i.e., IT application orchestration—conceptualized as dynamic capability—can enhance organizations' level of agility and competitive performance. Digital dynamic capability is tough to mimic and establish within organizations as it required specific, idiosyncratic, and heterogeneous competencies to develop [77, 78]. As such, this capability requires substantial undertakings toward embracing new digital technologies [39, 40].

## Dynamic capabilities and knowledge processes

As its definition and conceptualizations suggest, digital dynamic capability builds upon a rich foundation of the dynamic capabilities view (DCV) [78-82]. The DCV is a foundational strategic framework within the management and IS field [83, 84] and is built from a multiplicity of theoretical roots [85]. The DCV claims that under conditions of high economic turbulence, traditional resource-based capabilities do not provide organizations with a competitive edge [86-88]. Instead, within this framework, organizations seek a balance between strategies to remain stable in the process of delivering current business services distinctively and mobile so that they can anticipate and effectively address market disruptions and business changes [88]. The DCV, therefore, goes beyond existing resource-based perspectives principles and attempts to explain how organizations can obtain and *maintain* a competitive edge in turbulent economic environments [78] as traditional 'resource-based' capabilities seem to erode and cease to provide competitive gains under these particular business conditions [86]. These dynamic capabilities have been defined and conceptualized as sets of measurable and identifiable routines that have been widely validated through empirical studies [82, 89, 90]. In general, these capabilities can be conceived as the organizations' routines to integrate, build, reconfigure, gain and release internal competences and resources to address changing market and business ecosystem demands [78, 80]. In short, these capabilities can represent an organization's ability 'to act' under changing circumstances [58, 91]; a first 'derivative' of traditional resource-based capabilities: the ability to contribute to maintaining a competitive edge continuously.

The pivotal role of knowledge as an organizational resource has been well documented in the extant literature [92, 93] and in manifestos [e.g., 94]. In essence, knowledge drives the process of organizational learning and, thereby, enhances the organization's ability to rightfully and adequately identify customer needs and valuable markets [92]. Prior studies also embraced this view and investigated the relationship between knowledge processes and customer relationships [95], customer response capabilities [92], as well as hospital process capabilities [54]. Although the knowledge-based view of organizations strongly builds upon the organizational learning theories and literature [93, 96], recent studies converged both strategic management streams toward the core idea of knowledge-related dynamic capabilities. Knowledge processes represent the crucial operations for the input of knowledge assets [97]. As such, they focus on generating, analyzing, and distributing customer information for strategy formulation and implementation [54, 92, 98].

Making this specific for the cure sector, knowledge processes in hospitals are crucial for patients care, as acquiring new medical knowledge and insights can greatly impact patients' treatment [41]. Knowledge processes foster clinicians and medical staff to exchange and share medical and patient knowledge and, as such, these processes can be regarded as an effective way to integrate medical knowledge, enhance knowledge flow and cultivate the use of evidence-based care that will likely have a positive impact on the quality of care [99, 100]. Furthermore, knowledge processes facilitate transforming clinical data into patient-related insights, thereby supporting clinicians within hospitals to make informed decisions concerning diagnosis and treatment [50, 52, 101]. Thus, these data-driven processes allow clinicians to improve the patient treatment process and medical quality services and be more agile in their work [52]. As conceived in this study, knowledge processes conceptualized as a dynamic capability are closely related to the concept of absorptive capacity [102, 103]. Absorptive capacity can be considered the ability to capture and exploit the value of new (external) information, and transform this information (or knowledge) into the firm's knowledge base, and apply this new knowledge through innovation and competitive actions [102, 104]. In sum, knowledge processes are a crucial productive resource of organizations, and they enhance other organizational capabilities based on the degree of knowledge reach and richness the organization can achieve [105, 106].

There have only been a handful of articles that embrace the dynamic capabilities view when investigating the use and deployment of IT in healthcare [107]. For instance, Singh et al. [108] investigated how a home health care provider actively responded to externally imposed challenges (regulatory and environmental turbulence) by transforming its post-acute care through enhanced nursing practices; engaging key stakeholders (i.e., patient, physicians, nurses, managers) and implementing IT-enabled innovations. In a similar vein, Pablo et al. [109] demonstrated how a healthcare organization used a dynamic capability approach to strive for continuous organizational performance enhancements and what the particular role of information sharing was in this process. Wu and Hu [54] examined the interaction between knowledge assets and capabilities and their respective impact on hospital processes and organizational performance. Kislove et al. [110] also embraced the dynamic capabilities view and argued that capacity-building, the process of developing practice-based skills, is crucial to enhance knowledge mobilization in healthcare. A few studies embrace the concept of absorptive capacity as a dynamic capability. Gopalakrishna-Remani et al. [111] unfold the essential role of absorptive capacity in the process of influencing hospital top management beliefs and participation and its impact on hospital-wide EMR adoption. Likewise, Faber et al. (2017) investigated the role of absorptive capacity—among other vital antecedents—on eHealth adoption, and contradictory found that it does not significantly influence eHealth adoption. There are no studies (empirical nor theoretical) that conceptualize hospital departments' ability to anticipate patient needs, i.e., patient agility, following the dynamic capabilities view.

## The concept of organizational and patient agility

The DCV argues that organizations can respond to changing conditions while simultaneously proactively enact influencing the environment. Organizational agility has been considered a critical capability for sustained organizational success under the DCV [88]. This particular capability has been defined and conceptualized in many ways and through various theoretical lenses in the IS literature [26, 106, 112]. For instance, Park et al. [113] ground their conceptualization and operationalization in the information-processing theory [114] and argue that information processing capabilities strengthen the organization's sense-response processes to adapt to changing environmental conditions. Lu and Ramamurthy [65] embrace a complementarity perspective and perceive agility as the organization's ability to seize market opportunities and operationally adjustment capacity. Chakravarty et al. [26] adopt a contingency factors perspective and operationalize the multidimensional concept of agility through the organization's ability to anticipate and proactively respond to market dynamics, i.e., entrepreneurial agility and the organization's ability to react to events without needing substantial strategic changes, i.e., adaptive agility. A multidimensional view is also adopted by [35], who likewise perceive organizational agility as a higher-order multidimensional dynamic capability that allows organizations to effectively and efficiently sense and respond to environmental conditions. Roberts and Grover [115] synthesized that, although there seems to be ambiguity in definitions as reflected by the concepts' operationalized capabilities, a set of high-level characteristics can be devised from the extant literature. Hence, to a certain degree, all studies show two high-level organizational routines: deliberately 'sensing' and 'responding' to business events in the process of capturing business and market opportunities. These two organizational capabilities are crucial for organizations' success [31]. Hence, our paper perceives patient agility as a higher-order manifested type of dynamic capability that allows hospital departments to adequately 'sense' and 'respond' to patient needs, demands, and opportunities within a turbulent and fast-paced hospital ecosystem context [43, 88, 115, 116].

By addressing these crucial questions, this paper contributes to the medical informatics and information systems (IS) literature by unfolding the mechanisms through which the dual capacity of IT exploration and IT exploitation simultaneously drives patient agility in hospital departments.

# Research model and hypotheses

IT ambidexterity as a core organizational IT resource is expected to enhance hospital departments' level of patient sensing and responding capability (both conceptualized as higher-order dynamic capability) through digital dynamic capability (as a lower-order technical dynamic capability) and knowledge processes. Figure 1 demonstrates the research model and the associated hypotheses that will be clarified below. For the sake of simplicity, the figure does not demonstrate included control variables.

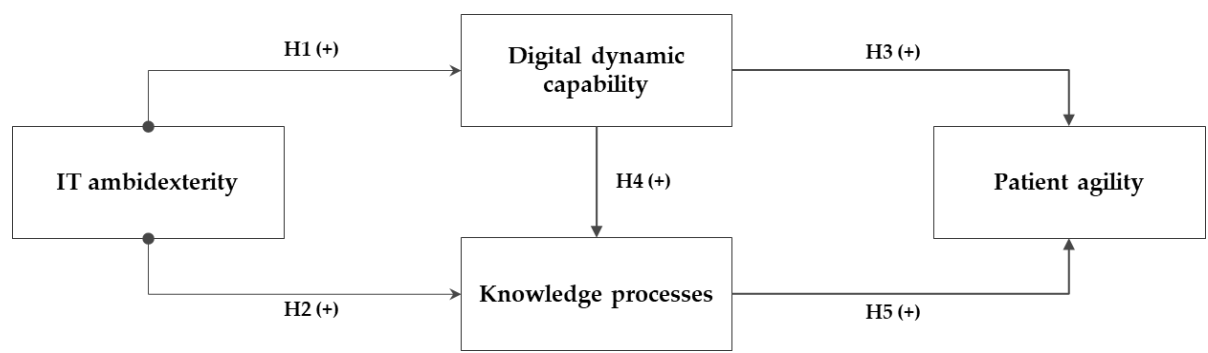

**Figure 1.** Research model

Investments in the organization's IT resources and assets are essential for the process of capability-building and gaining IT business value and a competitive edge [31, 106, 117-119]. IT resources are typically referred to by their aggregated latent components and qualities (e.g., hardware, software, networks, data sources) and IT-related managerial activities (e.g., IT planning, business connectivity) and how they related to business value [66, 120]. However, recent studies argue that IT-business value and organizational agility do not result from the deployment of isolated (non)IT resources and competencies. Instead, IT-business value seems to emerge from the complementarity to assimilate and re-align the IT resource portfolio to the changing business needs and demands [121].

IT can be a transformative force in hospitals and contribute to enhanced patient services, efficiency and effectiveness gains, and clinical care [10, 122]. However, IT implementations in hospitals are often exposed to cultural, organizational, and social challenges and inertia forces [10, 122, 123]. Therefore, an ambidextrous IT implementation strategy should be embraced, whereby short-term contributions (exploitation of current IT resources) and continuous progress of the IT resource portfolio (exploratory mode) drive IT-driven business transformation simultaneously [124]. When both short-term goals and ambitions are synchronized with the longer-term objectives, hospital departments' are better-equipped to develop digital capabilities, knowledge options and frame the hospital's business strategy and clinical practice [39, 106, 125].

IT exploration can be considered an enabler of digital dynamic capability. This mode promotes the use of and experimentation with new IT resources (e.g., new IT platform implementation, decision-support functionality, big data and clinical analytics, social media) as a basis to reshape existing patient services. On the other hand, IT exploitation focuses on using, enhancing, and repackaging existing IT resources (e.g., reuse or redesigning current EMR for new patient service development and ensuring hospital-wide accessibility to clinical patient data and information). Therefore, digital dynamic capability relates well to the dual capacity to aim for two disparate modes of operandi in managing the department's skills, qualities, and competencies to manage digital technologies and developments—like mobile, social media, big data analytics, robotic process automation, artificial intelligence (AI), cloud computing, Internet of Things (IoT)—for new patient service delivery. However, in practice, many organizations' struggle to reach IT ambidexterity results from resource constraints and conflicting ambitions and motives [126]. As the individual qualities of IT ambidexterity may, to some extent, strengthen hospital departments' digital options, they will likely not enhance the hospital department's digital dynamic capabilities in isolation [106]. The simultaneous engagement of IT exploration and IT exploitation will enhance the qualities and competencies to manage innovative digital technologies for new patient service development as they depend primarily on the organization's investment decision to deploy simultaneous short-term improvement activities and long-term innovations [127].

Thus, IT resources play a crucial role in acquiring, processing, organizing, and distributing knowledge and providing digital processes and knowledge options as enablers of agility [35, 106, 128]. This study argues that departments that can simultaneously engage in exploitation and exploration of its current IT resources portfolio will be better equipped to integrate existing and leverage new patient information sources, ensuring hospital-wide accessibility to clinical data and drive effective knowledge processes [129, 130]. By leveraging the two modes of IT management practices, hospital departments can effectively integrate and analyze patient knowledge, use it for interdepartmental meetings, and identify patients' needs for new health service development.

In line with this reasoning, this study now defines the following hypotheses:

**Hypothesis 1:** *The greater the hospital department's IT ambidexterity, the higher will be the degree of its digital dynamic capability.*

**Hypothesis 2:** *The greater the hospital department's IT ambidexterity, the more effective will be its knowledge processes.*

Digital dynamic capability is a crucial dynamic capability necessary to innovate and enhance business operations [39, 60, 131, 132]. Various prior studies investigated the benefits that result from developing a digital dynamic capability. Wang et al. [133] argue that digital dynamic capability allows leveraging IT and knowledge resources to deliver innovative services that customers value and contribute to organizational benefits. Coombs and Bierly [134] empirically showed that a sophisticated digital dynamic capability enables competitive advantages. Thus, the extant literature shows that digital dynamic capability drives organizations' ability to learn from experience in turbulent economic and competitive environments actively. Hence, in such an environment, it is essential to search continuously, identify, and absorb new knowledge and technological innovation such that they can be used to respond to changing customer behavior, demands, and wishes timely, adequately, and innovatively [28, 131]. These claims are likewise consistent with results from Westerman et al. (2012), Khin et al. [39], and Ritter and Pedersen [135], who showed that digital dynamic capability is crucial to deploy new innovative business models, enhance customer experiences, and improve business agility. By actively managing the opportunities provided by innovative technologies and responding to digital transformation, organizations can succeed in their digital options, products, and services [39].

A technological-driven capability is crucial for hospital departments that want to strive for patient agility in clinical practice because the process of achieving new digital patient service solutions is exceedingly dependent on its ability to manage digital technologies [39]. It requires proactively responding to digital transformation, mastering the state-of-the-art digital technologies, and deliberately developing innovative patient services using digital technology. Such a capability goes well beyond the notion of IT capabilities, i.e., aggregation of IT resources and IT competencies in the vast majority of empirical studies [66, 67, 136]. The development of a digital dynamic capability is tough to mimic and establish within hospital departments as it required specific, idiosyncratic, and heterogeneous competencies to develop [77, 78].

The digital dynamic capability allows hospital departments, e.g., to absorb and process sensitive patient information better, support clinicians in their decision-making processes, exchange clinical data, and facilitate patient health data accessibility [43, 137]. As such, developing this capability makes the department more receptive to new patient data and information. The accumulation and storing of knowledge necessary to develop these new technologies also improve a firm's ability to engage in transformation processes through its evaluation, use, and implementation. Finally, as a firm engages more in developing and mastering new technologies, they become more efficient in deploying the existing knowledge and, thus, generate more exploitative activities [138].

Hence, hospitals that actively invest and develop such a capability are likely to anticipate their patients' needs (of which they might be physically and mentally unaware) and respond fast to changes in the patient's health service needs using digital innovations and assessments of clinical outcomes [39, 115, 116]. Therefore, such a strategically significant capability is crucial for the departments' focus on quality, efficiency, essential patient information, and enhancing the patient's clinical journey. Based on the arguments given above and building upon the DCV, the following two hypotheses are defined:

**Hypothesis 3:** *The more developed the hospital department's digital dynamic capability, the higher the hospital department's patient agility.*

**Hypothesis 4:** *The more developed the hospital department's digital dynamic capability, the more effective will be the hospital department's knowledge processes.*

Previous scholarship demonstrated that knowledge-based capabilities and agility are two crucial capabilities that mediate the impact of IT resources and capabilities on business benefits [105, 106]. In the context of hospital departments, substantial investments in processing and analyzing patient data and information and adequate interdepartmental knowledge and information flow will drive the department's ability to anticipate the patients' current and future needs [92]. In clinical practice, the diagnosis and treatment processes are composed of a multitude of interactions and coordination between care activities in different activity levels and multiple types of knowledge [52]. Moreover, departments that are more aware of their patient needs through information knowledge processes are likely to harness new patient knowledge more effectively, make better clinical practice decisions, and support the treatment process [52, 92, 105]. Thus, through knowledge processes, the department can develop and redesign its core processes and capabilities. Mature knowledge-based processes drive transfer of knowledge across and within the department, uniquely deploy knowledge resources, and allow hospitals department to enhance business processes, and services, and better sense and seize business and patient services opportunities that ultimately can enhance business performance [54, 88, 91]. Recently, scholars showed that data and knowledge-driven capabilities, as intermediate constructs, contribute to hospital performance enhancements [139, 140]. Moreover, in hospital departments, patient agility as a crucial capability describes the competence of the medical professionals' ability to create patient value and drive patient satisfaction in a way that uniquely utilizes knowledge resources and processes [46, 54].

In sum, this study argues that knowledge processes are crucial in the process of reconfiguring its existing patient sensing and responding capabilities [103] and that these capabilities, to a great extent, rely on the integration of knowledge processes in the department [54, 93, 130]. Hence, this study defines the following:

**Hypothesis 5:** *The more effective the hospital department's knowledge processes, the higher the hospital department's patient agility.*

# Methods

A deductive and quantitative approach was used to address the study objectives. Hence, hypothesized relationships among key constructs are analyzed by first cross-sectionally collecting field data and then analyzing the obtaining survey data.

## Data collection procedure

An online survey was developed to capture clinicians' attitudes toward IT ambidexterity, digital dynamic capability, knowledge processes, and patient agility of their respective hospital departments. Hence, we adopt a practitioner-based approach that used subjective measures because hospitals are typically more willing to provide subjective data than sensible objective performance metrics, see for example, [56, 57]. In practice, perceptual measures on processes and practices positively correlate with objective data [141].

This survey was pretested on multiple occasions by five Master students and six medical practitioners and scholars to improve both the content and face validity of the survey items. The medical practitioners all had sufficient knowledge and experience to assess the survey items effectively to provide valuable improvement suggestions. The target population was (clinical) department heads- and managers, team-leads, and doctors under the assumption that, at the hospital department level, these respondents are actively involved in contact with patients or at least have an intelligible insight into the department's patient interactions, and the use of IT. Moreover, these are the foremost respondents who can provide insights into the unique and sometimes complicated situations where medical knowledge is exploited, enabling a unique treatment course [54]. This approach is a similar approach taken by many other key publications in the field surveying clinicians to obtain insights into how patients-based information affects the diagnosis, therapy, patient safety, and overall clinical practice and care to patients [54, 142-144]. Therefore, these respondents were considered to be the most important subject in this survey. Our single informant strategy is consistent with prior literature on specialized, not diversified units and departments [145].

Data were conveniently and anonymously collected between November 10$^{th,}$ 2019, and January 5$^{th,}$ 2020, sampled from Dutch hospitals through the five Master students' professional networks within Dutch hospitals.

Within the survey, comprehensible construct definitions were provided, and the survey followed a logical structure. In one of the final questions, the respondents were asked if they wanted to receive critical insights from the study. During the data collection process, various controls were also built so that each department completed the survey only once. Also, the respondents could withdraw their entries if they wanted to. The survey software registered 230 active and unique respondents. However, in total, 101 cases had to be removed because of unreliable data entries or no entries at all. Also, one respondent (administrative function) did not belong to our target population and had to be removed from the sample. In a final step, 21 additional respondents had to be removed due to substantial missing values (i.e., more than 15%). This study uses 107 complete survey responses for final analyses. Within the obtained sample, 36 respondents work for a University medical center (33.6%), 41 work for a specialized top clinical (training) hospital (38.3%), and the final 30 work for general hospitals (28%). Table 1 shows the demographics of the participating hospital departments. See also Appendix A for an overview of the survey responses per medical department.

This study accounts for possible non-response bias by using a *t*-test to assess whether or not there is a significant difference between the early respondents (*N*=66) and the late subsample (*N*=41 respondents) on the responses on the Likert scale questions. This assessment is crucial as non-response bias can significantly impact the study outcomes and requires careful examination [146, 147]. Hence, this study included various elements, including department age, the number of patients, and all construct items in the assessment. After running the analyses and assessing Levine's equality test (of variances) and the t-test for equality of mean-values, no significant difference could be detected. Hence, this confirms the absence of non-response bias. Finally, per suggestions of [148, 149], Harman's single-factor analysis was applied using exploratory factor analysis (in using IBM SPSS Statistics™ v24) to restrain possible common method bias [148, 149]. Hence, the current study sample is not affected by method biases, as no single factor is attributed to the majority of the variance.

## Constructs and items

The selection of constructs and measures was made following previous empirically validated work. Also, this study includes only measures that were suitable for departmental-level analyses. Since this research was done in a healthcare setting, some original items had to be slightly re-worded to fit the particular context. *IT ambidexterity* is operationalized using the item-level interaction terms of IT exploration and IT exploitation [35, 72]. Items were adopted from [35]. This study uses three measures from [39] for *digital dynamic capability* to represent the department's capability to manage innovative digital technologies for new patient service development. *Patient knowledge processes* refer to critical activities within the department that focus on generating, analyzing, and distributing patient-related information for strategy formulation and implementation. Six items based on the work of [92] are adopted. *Patient agility* concerns the departments' ability to sense and respond to patient needs adequately and is modeled as a higher-order (2$^{nd}$) dynamic capability comprising the first-order dimensions '*patient sensing capability*' and '*patient responding capability*' [28, 31, 106]. Hence, this study uses ten empirically validated measures from Roberts and Grover [28]. See Appendix C for a complete overview of the construct and their associated items with their respective item-to-construct loadings ($\lambda$), mean values ($\mu$), and the standard deviations (*Std*.). All of the above items were measured using a seven-point Likert scale.

This study controlled the outcomes for both 'size,' measured as full-time employees (log-normally distributed), and 'age' of the department (5-point Likert scale 1: 0–5years; 5: 25+ years).

**Table 1.** Demographics of participating hospital departments

| Element | Category | Frequency | Percentage |
|---|---|---|---|
| Hospital type | University medical center | 36 | 33.6% |
|  | Top clinical training hospital | 41 | 38.3% |
|  | General Hospital | 30 | 28% |
| Department age | 0–5 years | 28 | 26,2% |
|  | 6–10 years | 20 | 18,7% |
|  | 11–20 years | 20 | 18,7% |
|  | 20–25 years | 8 | 7,5% |
|  | Over 25 years | 31 | 29,0% |
| Experience at this particular department | 0–5 years | 49 | 45,8% |
|  | 6–10 years | 18 | 16,8% |
|  | 11–20 years | 28 | 26,2% |
|  | 20–25 years | 6 | 5,6% |
|  | Over 25 years | 6 | 5,6% |
| Amount of patients | < 4000 | 25 | 23,4% |
|  | 4000 – 6500 | 21 | 19,6% |
|  | 6500 – 9000 | 12 | 11,2% |
|  | 9000 – 11500 | 12 | 11,2% |
|  | 11500 – 14000 | 11 | 10,3% |
|  | > 14000 | 26 | 24,3% |

# Analyses and results

## Model estimation procedure and sample justification

The research model's hypothesized relationships are tested using Partial Least Squares (PLS) structural equation modeling (SEM). To estimate and model parameters SmartPLS version 3.2.9. is used [150]. In essence, PLS-SEM allows assessing both the measurement model, i.e., outer model [151], and the structural model (i.e., inner

9/23

model) of the research model, so that hypotheses can be tested [152]. The PLS algorithm establishes latent constructs from the factor scores. It, thereby, seemingly avoids factor indeterminacy [153], so that these scores then be applied in the following analyses [154]. A fundamental justification for using PLS-SEM is that its usage is appropriate in exploratory contexts and for the objective of theory development [153]. In this research, the focus is on prediction as to which the PLS algorithm assesses the explained variance ($R^2$) for all dependent constructs [153]. Also, PLS is less strict in terms of particular data distributions [151]. Another reason to justify the variance-based approach is that the current sample is relatively small [155]. The sample size does exceed minimum threshold values to obtain stable PLS outcomes [156]. A power analysis was done using G*Power [157]. Hence, this study assumes the conventional 80% statistical power and a 5% probability of error as input parameters, while the maximum number of predictors in the research model is three (when including the non-hypothesized direct effect of IT ambidexterity on agility). Based on G* Power's output parameters, a minimum sample of 38 cases were needed to detect an $R^2$ of at least 24%. The current sample of 107 far exceeds this minimum requirement. The estimation procedure makes use of the general recommended path weighting scheme algorithm [150].

## Assessment of the measurement model

Various analyses were done to determine the reliability and validity of the study constructs. In the first step, the internal consistency reliability is investigated using both the Cronbach's alpha measure (CA) and the composite reliability estimation (CR) value. In a subsequent step, this study assessed the convergent validity—using the average variance extracted (AVE)—of the first-order latent constructs [150]. All the AVE-values exceeded the lowest recommended mark of 0.50 [158]. Construct-to-item loadings were likewise investigated to determine the degree to which a variable contributes to explaining the variance of a particular construct while considering the other measurements. These loading also exceeded minimum thresholds. In a final step, discriminant validity was established through the assessment of three tests. First, cross-loadings were investigated [159]. Analyses show that all items load more strongly on their intended latent constructs than they correlate on other constructs. See also Appendix B. Secondly, the well-known Fornell-Larcker criterion is used [158]. In doing so, the square root of the AVE (see the diagonal entries in bold in Table 2) is compared with cross-correlation values. With this, each square root value should be larger than the cross-correlations [152]. As can be gleaned from Table 2, all Fornell-Larcker values are higher than the shared variances of the constructs with other constructs in the model. A final step in assessing discriminant validity is the newly developed heterotrait-monotrait ratio of correlations (HTMT) [160]. In general, acceptable outcomes of this analysis are HTMT-values that are below 0.85 (upper bound). Then, discriminant validity is established between constructs. The HTMT analyses show that all values are well below the threshold value of 0.85. Table 2 summarizes the entire assessment. The higher-order (formative) construct of patient agility was assessed using variance inflation factors (VIFs) values for the constructs patient sensing and patient responding capability. These VIF-values were well the conservative threshold of 3.5. Hence, no no multicollinearity is present within the research model [161].

As the reliability and validity of the model are now established, the model's associated fit indices can be assessed as well as the hypothesized relationships using the structural model.

**Table 2.** Convergent and discriminant validity assessment

|  | AVE | CA | CR | EXPLR | EXPLO | DDC | PSC | PRC | KP |
|---|---|---|---|---|---|---|---|---|---|
| *Constructs* | | | | | | | | | |
| EXPLR | 0.888 | 0.867 | 0.919 | **0.942** | | | | | |
| EXPLO | 0.790 | 0.937 | 0.960 | 0.502 | **0.889** | | | | |
| DDC | 0.783 | 0.862 | 0.916 | 0.584 | 0.631 | **0.885** | | | |
| PSC | 0.723 | 0.904 | 0.929 | 0.375 | 0.502 | 0.588 | **0.850** | | |
| PRC | 0.792 | 0.934 | 0.950 | 0.313 | 0.341 | 0.452 | 0.508 | **0.890** | |
| KP | 0.616 | 0.875 | 0.906 | 0.463 | 0.512 | 0.552 | 0.713 | 0.393 | **0.785** |

*Note: EXPLR: IT exploration EXPLO: IT exploitation; DDC: digital dynamic capability; PSC: patient sensing capability; PRC: patient responding capability; KP: patient knowledge processes*

## Model fit assessments

This study employs three metrics, i.e., (1) Standardized Root Mean Square Residual (SRMR)[2], (2) Stone–Geisser's test, and (3) the variance explained by the model ($R^2$) to assess the goodness-of-fit of the model. *First*, the newly developed SRMR metric is calculated. The SRMR metric calculates the difference between observed correlations and the model's implied correlations matrix [152, 162]. The obtained SRMR of 0.059 is well below the conservative threshold mark of 0.08, as proposed by [162].

*Secondly*, the Stone–Geisser's test ($Q^2$) is calculated using the blindfolding procedure to assess the model's predictive relevance. Hence, the current model's $Q^2$ values (for endogenous constructs*)* all far exceed 0, thereby indicating the overall model's predictive relevance.

*Finally*, $R^2$ values are analyzed. The structural model explains 47% of the variance for digital dynamic capability. ($R^2$=.47). The explained variance for patient knowledge processes is 36% and for patient agility 51%. These $R^2$ outcomes are considered moderate to substantial effects [163]. Based on the assessed four metrics, it can be concluded that the research model performs well compared with the base values and that sufficient model fit is obtained to test the hypotheses.

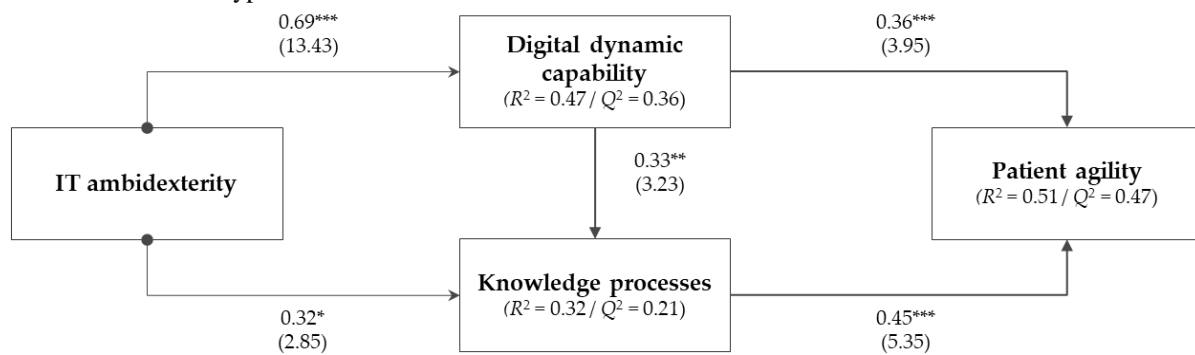

*$p \leq 0.05$, ** $p \leq 0.01$, *** $p \leq 0.001$

**Figure 2.** Structural model results

## Assessment of the structural model and hypotheses testing

Based on the outcomes of the non-parametric bootstrap resampling procedure [152], this study found support for the first hypothesis, i.e., IT ambidexterity positively impacts digital dynamic capability ($β = 0.69$; $t = 13.43$; $p < 0.0001$). Likewise, this study found support for hypothesis 2, i.e., IT ambidexterity → knowledge processes ($β = 0.32$, $t = 2.85$, $p = 0.0045$). Digital dynamic capability is positively associated with patient agility ($β = 0.36$; $t = 3.95$; $p = 0.0001$), providing support for hypotheses 3. Also, the structural model results support hypothesis 4, i.e., digital dynamic capability → knowledge processes ($β = 0.33$, $t = 3.23$, $p = 0.0012$). Hence, outcomes show that digital dynamic capability 'partially' mediates the effect of IT ambidexterity on knowledge processes [152, 164]. Finally, the results support hypothesis 5, i.e., knowledge processes are positively associated with patient agility ($β = 0.45$, $t = 5.35$ $p < 0.0001$). Thus, it can be concluded that also, partial mediation characterizes the triangular relationship between digital dynamic capability, knowledge processes, and patient agility.

The bootstrapped PLS results showed non-significant effects for the included control variables: 'size' ($β = -0.10$, $t = 0.79$, $p = 0.86$), 'age' ($β = -0.01$, $t = 0.17$, $p = 0.43$). Figure 2 summarizes the structural model assessment results.

---

[2] Note, however, that the first two metrics for model fit should be interpreted with caution as these metrics are not fully established PLS-SEM evaluation criteria.

# Discussion

The digital transformation brings about an unprecedented challenge for modern-day hospitals [165, 166]. Decision-makers and stakeholders across the hospital need to make sure that digital resources and technological innovations are aligned and deployed with care to enhance efficiencies, decision-making, and quality of services, so that personalized and patient-centered care can be delivered [167]. Thus, it is needless to say that digital innovations can improve existing processes and medical procedures for diagnostics and patient treatment. The current study makes substantial theoretical and practical contributions, which will be discussed next.

## Implications for theory and practice

The process of digitizing existing patient services and come up with new digital solutions remains time-consuming and challenging in many ways. Also, from a research perspective, there is still a limited understanding of how IT resources and the digital capability-building processes can facilitate patient agility and contribute to the much-needed insights on obtaining value from IT at the departmental level [35, 168, 169]. This study aimed at addressing these particular gaps in the literature. Notably, this study designed and tested a research model, using a sample of 107 hospital departments from the Netherlands, arguing that IT ambidexterity would drive a department's patient agility by first enabling digital dynamic capability and the department's knowledge processes. Outcomes of this study found support for these foundational claims. This study's structural model analyses unfolded that IT ambidexterity is a crucial antecedent of digital dynamic capability and knowledge processes. These crucial capabilities and processes, in turn, significantly impact the departments' ability to adequately sense and respond to patient needs and wishes, i.e., patient agility.

Evidence unfolds that digital dynamic capability partially mediates the relationship between IT ambidexterity and knowledge processes. Similarly, a partial mediation characterizes the triangular relationship between digital dynamic capability, knowledge processes, and patient agility. These outcomes corroborate existing IT-enabled agility and dynamic capability studies [105, 108, 115, 170]. The results also support the core idea that the hospital department's capacity to obtain value from its knowledge assets is a crucial success factor in achieving patient agility [128, 170].

This study embraces the dynamic capabilities, and knowledge-based view when it comes to IT resources deployments and advances the current insights on the resource and capability-building perspective [65, 78, 106, 168]. It does so by unfolding the nomological path from 'resources' to the 'IT-enabled value-perspective.' [21]

Outcomes of this study suggest that hospitals—that are committed to the process of ambidextrously managing their IT resources—are more proficient in promptly sensing and responding to patients' medical needs and demands. These theoretical contributions are valuable as these particular insights remained unclear in the extant literature, and future research can take these particular insights into account when investigating the IT benefits in hospitals. Likewise, unfolding the benefits of hospital departments' dual capacity to aim for two disparate things at the same time using empirical data is very relevant from a practical perspective as the business value of IT and the preceding IT investments can be justified [35, 171, 172]. The outcomes corroborate with the "theory of swift and even patient flow" [44], in that digital capabilities support the process of optimizing current hospital assets and help adequately responding to patient's needs by improving hospital operations (e.g., better diagnoses, scheduling, and coordination of patient care). Hence, it supports the 'call' for researchers to demonstrate the best ways of optimizing digital solutions in health care [21].

This study provides hospital department managers and decision-makers with valuable practical implications. Hospital departments must direct IT investments to bring about the highest IT business value, given the many substantial challenges to ensure high quality across the patient care delivery continuum. This research shows that IT ambidextrous departments can adequately develop new innovative digital opportunities and patient services to enhance the hospital department's knowledge processes and patient agility levels. This development path is crucial for successful hospital departments that strive to enhance the patient's clinical journey and provide patients with fitting health services. However, it is important to note that IT ambidexterity can help hospital departments indirectly obtain high levels of patient agility. However, this development might be hindered if departments do not fully leverage their dual capacity of IT exploration and IT exploitation to drive digital dynamic capability and knowledge processes and enhance their patient agility.

Digital dynamic capability is crucial in the development of knowledge processes and patient agility. Hospital department managers should develop the core competencies, knowledge, and skills to process patient information better, adequately respond to digital transformation, master the state-of-the-art digital technologies, and deliberately develop innovative patient services using digital technology. Hospital department managers should also be aware of the crucial role of knowledge processes. Mature knowledge processes enhance decision-making processes and drive patient agility in hospital departments. Therefore, they should dedicate their resources to leverage these capabilities capability fully so that they are better equipped to search, identify, and absorb new technological innovations, integrate, process, and exchange patient information and use them for decision-making processes and to anticipate and respond fast to changes in the patient's health service needs. Our study results highlight the need for hospital departments to focus more on patient agility, a crucial antecedent of enhanced patient care. Hospital department managers and decision-makers should also deliberately pay attention to end-user's psychological meaningfulness, stakeholder involvement, and providing adequate resourcing and infrastructures when implementing new digital technologies [33, 173-175]. These aspects are crucial when implementing new digital technologies so that the hospital staff is supported and perceived value can be related to individual behavior changes and key stakeholders' needs and expectations. The outcomes are particularly relevant for practitioners now, as hospitals worldwide need to take action to transform healthcare delivery processes using digital technologies and increase clinical productivity during the COVID-19 crisis [176].

In summary, hospital departments should strive to be agile in the modern turbulent economic environment. This study provides crucial insights and guidance to achieve this.

## Limitations and future studies

Several study limitations should be mentioned. These limitations suggest future research avenues. This study used self-reported data to test the developed hypotheses, as obtaining objective measures is typically a challenging endeavor [177-180]. The decision to use self-reported data is still justifiable, as empirical outcomes as these data types are strongly correlated to objective measures [141, 179, 181]. Another concern is that data were collected using the single informant strategy. As such, method bias could still be a concern. This study did pay considerable attention to account for possible measurement errors and method bias. Future research could embrace a matched-pair design where different respondents address independent (explanatory) and dependent constructs. Another opportunity for future research is triangulating the included measures with, e.g., potentially available archival data from public sources. These insights, next to possible applying the current model to other countries, could help validate the outcomes further. Also, a more substantial sample of hospital departments will further contribute to the robustness of the results. Scholars could confirm this research's outcomes using a replication study in different (non-Western) countries. Future research could also investigate patient agility, focusing on specific departments as the current study encompasses various participating departments. Focusing on a few departments with more responses could capture a richer view of the subject matter.

Finally, the current study did not include patient service performance outcomes and benefits beyond our paper's current scope. Hence, it would be interesting to investigate the relationships between patient agility and the hospital department's performance outcomes as patient agility is considered a crucial ingredient in delivering high-quality patient value and overall streamlined patient journeys. Hence, this research's outcomes inform further research about whether patient agility impacts clinical care quality and efficacy. Scholars could then investigate patient agility's contribution to increasing, e.g., clinical productivity and quality enhancement during different stages of the COVID-19 pandemic [176]. Finally, future work could also involve the patient engagement and digital technology co-design perspectives [175, 182, 183].

*Declarations of interest*

None.

*Acknowledgments*

We want to thank Josja Willems, Reinier Dickhout, Rick Smulders, Yves-Sean Mahamit, and Renaldo Kalicharan for their valuable contributions to the data collection and for sharing their perspectives in numerous discussions. Also, we would like to express our gratitude to all participating hospitals. Your active role made this a success.

# Appendix A: Survey response per medical department

| Department | # responses | % of total |
|---|---|---|
| General Internal Medicine | 3 | 3% |
| Anesthesiology | 4 | 4% |
| Pharmacy | 2 | 2% |
| Cardiology | 7 | 7% |
| Cardiothoracic surgery | 2 | 2% |
| Surgery | 8 | 7% |
| Dermatology | 3 | 3% |
| Endocrinology | 1 | 1% |
| Geriatrics | 1 | 1% |
| Infectious diseases | 1 | 1% |
| Intensive Care Adults | 5 | 5% |
| Pediatrics | 8 | 7% |
| Neonatology | 2 | 2% |
| Clinical immunology & Rheumatology | 2 | 2% |
| Clinical Oncology | 2 | 2% |
| Lung diseases | 2 | 2% |
| Gastrointestinal and liver diseases | 4 | 4% |
| Neurosurgery | 2 | 2% |
| Neurology | 4 | 4% |
| Kidney diseases | 3 | 3% |
| Ophthalmology | 2 | 2% |
| Orthopedics | 5 | 5% |
| Psychiatry | 2 | 2% |
| Revalidation | 1 | 1% |
| First aid | 6 | 6% |
| Urology | 3 | 3% |
| Vascular medicine | 2 | 2% |
| Obstetrics / Gynecology | 9 | 8% |
| Medical imaging | 6 | 6% |
| Day treatment | 3 | 3% |
| Radiotherapy | 1 | 1% |
| Paramedic care | 1 | 1% |
| **Total** | **107** | **100%** |

# Appendix B: Cross-loading analysis for the first-order factors

|  | EXPLR | EXPLO | DDC | PSC | PSR | KP |
|---|---|---|---|---|---|---|
| Explore1 | **0.921** | 0.487 | 0.543 | 0.345 | 0.211 | 0.418 |
| Explore2 | **0.944** | 0.454 | 0.494 | 0.238 | 0.239 | 0.338 |
| Explore3 | **0.944** | 0.419 | 0.529 | 0.356 | 0.350 | 0.413 |
| Exploit1 | 0.514 | **0.860** | 0.578 | 0.403 | 0.379 | 0.442 |
| Exploit2 | 0.397 | **0.915** | 0.575 | 0.447 | 0.283 | 0.403 |
| Exploit3 | 0.374 | **0.889** | 0.513 | 0.440 | 0.214 | 0.420 |
| Dig1 | 0.441 | 0.529 | **0.886** | 0.436 | 0.422 | 0.432 |
| Dig2 | 0.547 | 0.526 | **0.895** | 0.506 | 0.411 | 0.486 |
| Dig3 | 0.480 | 0.595 | **0.856** | 0.525 | 0.441 | 0.467 |
| Sense1 | 0.381 | 0.481 | 0.507 | **0.884** | 0.473 | 0.645 |
| Sense2 | 0.484 | 0.512 | 0.502 | **0.760** | 0.346 | 0.560 |
| Sense3 | 0.261 | 0.357 | 0.463 | **0.893** | 0.552 | 0.593 |
| Sense4 | 0.113 | 0.409 | 0.455 | **0.791** | 0.372 | 0.545 |
| Sense5 | 0.191 | 0.302 | 0.432 | **0.868** | 0.452 | 0.560 |
| Respond1 | 0.216 | 0.299 | 0.405 | 0.481 | **0.935** | 0.294 |
| Respond2 | 0.239 | 0.300 | 0.442 | 0.436 | **0.918** | 0.240 |
| Respond3 | 0.310 | 0.312 | 0.412 | 0.444 | **0.917** | 0.321 |
| Respond4 | 0.292 | 0.356 | 0.503 | 0.498 | **0.868** | 0.352 |
| Respond5 | 0.227 | 0.230 | 0.416 | 0.520 | **0.865** | 0.500 |
| KP1 | 0.277 | 0.266 | 0.193 | 0.361 | 0.164 | **0.676** |
| KP2 | 0.337 | 0.372 | 0.481 | 0.596 | 0.339 | **0.810** |
| KP3 | 0.389 | 0.275 | 0.395 | 0.516 | 0.304 | **0.715** |
| KP4 | 0.314 | 0.412 | 0.412 | 0.461 | 0.099 | **0.790** |
| KP5 | 0.357 | 0.404 | 0.476 | 0.581 | 0.320 | **0.857** |

*Note: EXPLR: IT exploration EXPLO: IT exploitation; DDC: digital dynamic capability; PSC: patient sensing capability; PRC: patient responding capability; KP: knowledge processes*

# Appendix C: Survey constructs and items and descriptive statistics

| Construct | | Measurement item | λ | μ | Std. |
|---|---|---|---|---|---|
| IT ambidexterity | \multicolumn{5}{l}{*Please indicate the ability of your department to: (1. Strongly disagree–7. Strongly agree)*} |
| | \multicolumn{5}{l}{*IT exploration capability*} |
| | EXPLR1 | Acquire new IT resources (e.g., potential IT applications, critical IT skills) | 0.860 | 4.01 | 1.67 |
| | EXPLR2 | Experiment with new IT resources | 0.915 | 3.81 | 1.62 |
| | EXPLR3 | Experiment with new IT management practices | 0.889 | 3.43 | 1.62 |
| | \multicolumn{5}{l}{*IT exploitation capability*} |
| | EXPLO1 | Reuse existing IT components, such as hardware and network resources | 0.921 | 5.29 | 1.28 |
| | EXPLO2 | Reuse existing IT applications and services | 0.944 | 5.18 | 1.32 |
| | EXPLO3 | Reuse existing IT skills | 0.944 | 5.13 | 1.25 |
| Dig. dynamic capability | \multicolumn{5}{l}{*Please indicate the level of your department's capabilities in following areas (1. Strongly disagree–7. Strongly agree).*} |
| | DDC1 | Responding to digital transformation | 0.886 | 4.33 | 1.56 |
| | DDC2 | Mastering the state-of-the-art digital technologies | 0.895 | 3.69 | 1.48 |
| | DDC3 | Developing innovative patient services using digital technology | 0.856 | 4.74 | 1.63 |
| Patient agility | \multicolumn{5}{l}{*Indicate the degree to which you agree or disagree with the following statements about whether the department can (1 – strongly disagree 7 – strongly agree)*} |
| | \multicolumn{5}{l}{*Patient sensing capability*} |
| | SENSE1 | We continuously discover additional needs of our patients of which they are unaware | 0.884 | 4.09 | 1.66 |
| | SENSE2 | We extrapolate key trends for insights on what patients will need in the future | 0.760 | 4.43 | 1.63 |
| | SENSE3 | We continuously anticipate our patients' needs even before they are aware of them | 0.893 | 4.03 | 1.68 |
| | SENSE4 | We attempt to develop new ways of looking at patients and their needs | 0.791 | 4.72 | 1.52 |
| | SENSE5 | We sense our patient's needs even before they are aware of them | 0.868 | 3.94 | 1.66 |
| | \multicolumn{5}{l}{*Patient responding capability*} |
| | RESPOND1 | We respond rapidly if something important happens with regard to our patients | 0.935 | 4.76 | 1.71 |
| | RESPOND2 | We quickly implement our planned activities with regard to patients | 0.918 | 4.11 | 1.62 |
| | RESPOND3 | We quickly react to fundamental changes with regard to our patients | 0.917 | 4.54 | 1.53 |
| | RESPOND4 | When we identify a new patient need, we are quick to respond to it | 0.868 | 4.52 | 1.42 |
| | RESPOND5 | We are fast to respond to changes in our patient's health service needs | 0.865 | 4.52 | 1.50 |
| Knowledge processes | \multicolumn{5}{l}{*Indicate the degree to which you agree or disagree with the following statements about whether the department can (1 – strongly disagree 7 – strongly agree)*} |
| | KP1 | We regularly meet patients to learn about their current and potential needs for new health services | 0.676 | 3.99 | 1.64 |
| | KP2 | Our knowledge of patients' needs is thorough | 0.810 | 4.10 | 1.69 |
| | KP3 | We systematically process and analyze patient data and information | 0.715 | 4.59 | 1.67 |
| | KP4 | We regularly study our patient's needs for new health service development | 0.790 | 4.02 | 1.53 |
| | KP5 | We have interdepartmental meetings regularly to discuss patient's needs | 0.857 | 4.04 | 1.71 |
| | KP6 | Our department spend time discussing patient's future needs with other (clinical) departments | 0.838 | 3.97 | 1.65 |

## References

1. Kostopoulos, K.C., Y.E. Spanos, and G.P. Prastacos. *The resource-based view of the firm and innovation: identification of critical linkages*. in *The 2nd European Academy of Management Conference*. 2002.
2. Hitt, M.A., R.E. Hoskisson, and H. Kim, *International diversification: Effects on innovation and firm performance in product-diversified firms.* Academy of Management journal, 1997. **40**(4): p. 767-798.
3. Aral, S. and P. Weill, *IT assets, organizational capabilities, and firm performance: How resource allocations and organizational differences explain performance variation.* Organization Science, 2007. **18**(5): p. 763-780.
4. Joshi, K.D., et al., *Changing the competitive landscape: Continuous innovation through IT-enabled knowledge capabilities.* Information Systems Research, 2010. **21**(3): p. 472-495.
5. Forés, B. and C. Camisón, *Does incremental and radical innovation performance depend on different types of knowledge accumulation capabilities and organizational size?* Journal of Business Research, 2016. **69**(2): p. 831-848.
6. Feldman, S.S., S. Buchalter, and L.W. Hayes, *Health information technology in healthcare quality and patient safety: literature review.* JMIR medical informatics, 2018. **6**(2): p. e10264.
7. McCullough, J.S., et al., *The effect of health information technology on quality in US hospitals.* Health Affairs, 2010. **29**(4): p. 647-654.
8. Lenz, R. and M. Reichert, *IT support for healthcare processes–premises, challenges, perspectives.* Data & Knowledge Engineering, 2007. **61**(1): p. 39-58.
9. Girgis, A., et al., *eHealth system for collecting and utilizing patient reported outcome measures for personalized treatment and care (PROMPT-Care) among cancer patients: mixed methods approach to evaluate feasibility and acceptability.* Journal of medical Internet research, 2017. **19**(10): p. e330.
10. Kohli, R. and S.S.-L. Tan, *Electronic health records: how can IS researchers contribute to transforming healthcare?* Mis Quarterly, 2016. **40**(3): p. 553-573.
11. Sligo, J., et al., *A literature review for large-scale health information system project planning, implementation and evaluation.* International journal of medical informatics, 2017. **97**: p. 86-97.
12. Kruse, C.S. and A. Beane, *Health information technology continues to show positive effect on medical outcomes: systematic review.* Journal of medical Internet research, 2018. **20**(2): p. e41.
13. Elton, J. and A. O'Riordan, *Healthcare disrupted: Next generation business models and strategies*. 2016: John Wiley & Sons.
14. Zheng, X., et al., *Accelerating Health Data Sharing: A Solution Based on the Internet of Things and Distributed Ledger Technologies.* Journal of medical Internet research, 2019. **21**(6): p. e13583.
15. Kuo, M.-H., *Opportunities and challenges of cloud computing to improve health care services.* Journal of medical Internet research, 2011. **13**(3): p. e67.
16. Chen, P.-T., C.-L. Lin, and W.-N. Wu, *Big data management in healthcare: Adoption challenges and implications.* International Journal of Information Management, 2020. **53**: p. 102078.
17. Lin, Y.-K., M. Lin, and H. Chen, *Do Electronic Health Records Affect Quality of Care? Evidence from the HITECH Act.* Information Systems Research, 2019. **30**(1): p. 306-318.
18. Paré, G., M. Jaana, and C. Sicotte, *Systematic review of home telemonitoring for chronic diseases: the evidence base.* Journal of the American Medical Informatics Association, 2007. **14**(3): p. 269-277.
19. Chiasson, M., et al., *Expanding multi-disciplinary approaches to healthcare information technologies: what does information systems offer medical informatics?* International Journal of Medical Informatics, 2007. **76**: p. S89–S97.
20. Wang, Y., et al., *An integrated big data analytics-enabled transformation model: Application to health care.* Information & Management, 2018. **55**(1): p. 64-79.
21. Jones, S.S., et al., *Unraveling the IT productivity paradox—lessons for health care.* N Engl J Med, 2012. **366**(24): p. 2243-2245.
22. Chiasson, M.W. and E. Davidson, *Pushing the contextual envelope: developing and diffusing IS theory for health information systems research.* Information and Organization, 2004. **14**(3): p. 155-188.
23. Andargoli, A.E., et al., *Health information systems evaluation frameworks: a systematic review.* International journal of medical informatics, 2017. **97**: p. 195-209.
24. Nair, A. and D. Dreyfus, *Technology alignment in the presence of regulatory changes: The case of meaningful use of information technology in healthcare.* International journal of medical informatics, 2018. **110**: p. 42-51.
25. Mackert, M., et al., *Health literacy and health information technology adoption: the potential for a new digital divide.* Journal of medical Internet research, 2016. **18**(10): p. e264.
26. Chakravarty, A., R. Grewal, and V. Sambamurthy, *Information technology competencies, organizational agility, and firm performance: Enabling and facilitating roles.* Information systems research, 2013. **24**(4): p. 976-997.

27. Rai, A. and X. Tang, *Leveraging IT capabilities and competitive process capabilities for the management of interorganizational relationship portfolios.* Information Systems Research, 2010. **21**(3): p. 516-542.
28. Roberts, N. and V. Grover, *Leveraging information technology infrastructure to facilitate a firm's customer agility and competitive activity: An empirical investigation.* Journal of Management Information Systems, 2012. **28**(4): p. 231-270.
29. Fink, L., *How do IT capabilities create strategic value? Toward greater integration of insights from reductionistic and holistic approaches.* European Journal of Information Systems, 2011. **20**(1): p. 16-33.
30. Brynjolfsson, E. and L.M. Hitt, *Beyond computation: Information technology, organizational transformation and business performance.* The Journal of Economic Perspectives, 2000. **14**(4): p. 23-48.
31. Overby, E., A. Bharadwaj, and V. Sambamurthy, *Enterprise agility and the enabling role of information technology.* European Journal of Information Systems, 2006. **15**(2): p. 120-131.
32. Carr, N.G., *IT doesn't matter.* Educause Review, 2003. **38**: p. 24-38.
33. Kumar, M., et al., *"Context" in healthcare information technology resistance: A systematic review of extant literature and agenda for future research.* International Journal of Information Management, 2020. **51**: p. 102044.
34. Hessels, A., et al., *Impact of heath information technology on the quality of patient care.* On-line journal of nursing informatics, 2015. **19**.
35. Lee, O.-K., et al., *How does IT ambidexterity impact organizational agility?* Information Systems Research, 2015. **26**(2): p. 398-417.
36. de Guinea, A.O. and L. Raymond, *Enabling innovation in the face of uncertainty through IT ambidexterity: A fuzzy set qualitative comparative analysis of industrial service SMEs.* International Journal of Information Management, 2020. **50**: p. 244-260.
37. Gaughan, D., *Use Bimodal and Pace-Layered IT Together to Deliver Digital Business Transformation*, in *Gartner*. 2016, Gartner, Inc.
38. Mesaglio, M., S. Adnams, and S. Mingay, *Kick-Start Bimodal IT by Launching Mode 2*, in *Gartner*. 2015, Gartner, Inc.
39. Khin, S. and T.C. Ho, *Digital technology, digital capability and organizational performance: A mediating role of digital innovation.* International Journal of Innovation Science, 2019. **11**(2): p. 177-195.
40. Božič, K. and V. Dimovski, *Business intelligence and analytics use, innovation ambidexterity, and firm performance: A dynamic capabilities perspective.* The Journal of Strategic Information Systems, 2019. **28**(4): p. 101578.
41. Ryu, S., S.H. Ho, and I. Han, *Knowledge sharing behavior of physicians in hospitals.* Expert Systems with applications, 2003. **25**(1): p. 113-122.
42. Wu, H. and Z. Deng, *Knowledge collaboration among physicians in online health communities: A transactive memory perspective.* International Journal of Information Management, 2019. **49**: p. 13-33.
43. Van de Wetering, R. *IT ambidexterity and patient agility: the mediating role of digital dynamic capability. In Proceedings of the Twenty-Ninth European Conference on Information Systems (ECIS)*. 2021. Marrakesh, Morocco: AIS.
44. Devaraj, S., T.T. Ow, and R. Kohli, *Examining the impact of information technology and patient flow on healthcare performance: A Theory of Swift and Even Flow (TSEF) perspective.* Journal of Operations Management, 2013. **31**(4): p. 181-192.
45. Prgomet, M., A. Georgiou, and J.I. Westbrook, *The impact of mobile handheld technology on hospital physicians' work practices and patient care: a systematic review.* Journal of the American Medical Informatics Association, 2009. **16**(6): p. 792-801.
46. Bradley, R., et al. *An examination of the relationships among IT capability intentions, IT infrastructure integration and quality of care: A study in US hospitals*. in *2012 45th Hawaii International Conference on System Sciences*. 2012. IEEE.
47. Sim, I., *Mobile devices and health.* New England Journal of Medicine, 2019. **381**(10): p. 956-968.
48. Karaca, Y., et al., *Mobile cloud computing based stroke healthcare system.* International Journal of Information Management, 2019. **45**: p. 250-261.
49. Mosa, A.S.M., I. Yoo, and L. Sheets, *A systematic review of healthcare applications for smartphones.* BMC medical informatics and decision making, 2012. **12**(1): p. 67.
50. Van de Wetering, R., *IT-Enabled Clinical Decision Support: An Empirical Study on Antecedents and Mechanisms.* Journal of Healthcare Engineering, 2018. **2018**: p. 10.
51. Salge, T.O. and A. Vera, *Hospital innovativeness and organizational performance: Evidence from English public acute care.* Health Care Management Review, 2009. **34**(1): p. 54-67.

52. Li, W., et al., *Integrated clinical pathway management for medical quality improvement–based on a semiotically inspired systems architecture.* European Journal of Information Systems, 2014. **23**(4): p. 400-417.
53. Krasuska, M., et al., *Technological Capabilities to Assess Digital Excellence in Hospitals in High Performing Health Care Systems: International eDelphi Exercise.* Journal of Medical Internet Research, 2020. **22**(8): p. e17022.
54. Wu, L. and Y.-P. Hu, *Examining knowledge management enabled performance for hospital professionals: A dynamic capability view and the mediating role of process capability.* Journal of the Association for Information Systems, 2012. **13**(12): p. 976.
55. Fadlalla, A. and N. Wickramasinghe, *An integrative framework for HIPAA-compliant I* IQ healthcare information systems.* International Journal of Health Care Quality Assurance, 2004.
56. McCracken, M.J., T.F. McIlwain, and M.D. Fottler, *Measuring organizational performance in the hospital industry: an exploratory comparison of objective and subjective methods.* Health services management research, 2001. **14**(4): p. 211-219.
57. Devaraj, S. and R. Kohli, *Performance impacts of information technology: Is actual usage the missing link?* Management science, 2003. **49**(3): p. 273-289.
58. Winter, S.G., *Understanding dynamic capabilities.* Strategic management journal, 2003. **24**(10): p. 991-995.
59. Danneels, E., *The dynamics of product innovation and firm competences.* Strategic management journal, 2002. **23**(12): p. 1095-1121.
60. Li, T.C. and Y.E. Chan, *Dynamic information technology capability: Concept definition and framework development.* The Journal of Strategic Information Systems, 2019. **28**(4): p. 101575.
61. Wang, C.L. and P.K. Ahmed, *Dynamic capabilities: A review and research agenda.* International journal of management reviews, 2007. **9**(1): p. 31-51.
62. Zheng, S., W. Zhang, and J. Du, *Knowledge-based dynamic capabilities and innovation in networked environments.* Journal of knowledge management, 2011.
63. Chiasson, M.W. and E. Davidson, *Taking industry seriously in information systems research.* Mis Quarterly, 2005: p. 591-605.
64. Liedtka, J.M., *Formulating hospital strategy: moving beyond a market mentality.* Health Care Management Review, 1992(17): p. 21–26.
65. Lu, Y. and K. Ramamurthy, *Understanding the link between information technology capability and organizational agility: An empirical examination.* 2011. **35**(4): p. 931-954.
66. Kim, G., et al., *IT capabilities, process-oriented dynamic capabilities, and firm financial performance.* Journal of the Association for Information Systems, 2011. **12**(7): p. 487.
67. Chen, Y., et al., *IT capability and organizational performance: the roles of business process agility and environmental factors.* European Journal of Information Systems, 2014. **23**(3): p. 326-342.
68. Wiesböck, F. and T. Hess, *Digital innovations.* Electronic Markets, 2020. **30**(1): p. 75-86.
69. March, J.G., *Exploration and exploitation in organizational learning.* Organization science, 1991. **2**(1): p. 71-87.
70. Raisch, S., et al., *Organizational ambidexterity: Balancing exploitation and exploration for sustained performance.* Organization science, 2009. **20**(4): p. 685-695.
71. Jansen, J.J., F.A. Van Den Bosch, and H.W. Volberda, *Exploratory innovation, exploitative innovation, and performance: Effects of organizational antecedents and environmental moderators.* Management science, 2006. **52**(11): p. 1661-1674.
72. Gibson, C.B. and J. Birkinshaw, *The antecedents, consequences, and mediating role of organizational ambidexterity.* Academy of management Journal, 2004. **47**(2): p. 209-226.
73. Tushman, M.L. and C.A. O'Reilly III, *Ambidextrous organizations: Managing evolutionary and revolutionary change.* California management review, 1996. **38**(4): p. 8-29.
74. Benitez, J., G. Ray, and J. Henseler, *Impact of information technology infrastructure flexibility on mergers and acquisitions.* MIS Quarterly, 2018. **42**(1).
75. Vitari, C., et al., *Antecedents of IT dynamic capabilities in the context of the digital data genesis.* 2012.
76. Queiroz, M., et al., *The role of IT application orchestration capability in improving agility and performance.* The Journal of Strategic Information Systems, 2018. **27**(1): p. 4-21.
77. Tripsas, M., *Surviving radical technological change through dynamic capability: Evidence from the typesetter industry.* Industrial and corporate Change, 1997. **6**(2): p. 341-377.
78. Teece, D.J., G. Pisano, and A. Shuen, *Dynamic capabilities and strategic management.* Strategic management journal, 1997. **18**(7): p. 509-533.
79. Teece, D.J., *Explicating dynamic capabilities: the nature and microfoundations of (sustainable) enterprise performance.* Strategic management journal, 2007. **28**(13): p. 1319-1350.

80. Eisenhardt, K.M. and J.A. Martin, *Dynamic capabilities: what are they?* Strategic management journal, 2000. **21**(10-11): p. 1105-1121.
81. Van de Wetering, R. *Enterprise Architecture Resources, Dynamic Capabilities, and their Pathways to Operational Value. In Proceedings of the Fortieth International Conference on Information Systems*. 2019. Munich: AIS.
82. Mikalef, P., A. Pateli, and R. van de Wetering, *IT architecture flexibility and IT governance decentralisation as drivers of IT-enabled dynamic capabilities and competitive performance: The moderating effect of the external environment.* European Journal of Information Systems, 2020: p. 1-29.
83. Schilke, O., *On the contingent value of dynamic capabilities for competitive advantage: The nonlinear moderating effect of environmental dynamism.* Strategic Management Journal, 2014. **35**(2): p. 179-203.
84. Pavlou, P.A. and O.A. El Sawy, *Understanding the elusive black box of dynamic capabilities.* Decision Sciences, 2011. **42**(1): p. 239-273.
85. Di Stefano, G., M. Peteraf, and G. Verona, *The organizational drivetrain: A road to integration of dynamic capabilities research.* Academy of Management Perspectives, 2014. **28**(4): p. 307-327.
86. Drnevich, P.L. and A.P. Kriauciunas, *Clarifying the conditions and limits of the contributions of ordinary and dynamic capabilities to relative firm performance.* Strategic Management Journal, 2011. **32**(3): p. 254-279.
87. Wilden, R. and S.P. Gudergan, *The impact of dynamic capabilities on operational marketing and technological capabilities: investigating the role of environmental turbulence.* Journal of the Academy of Marketing Science, 2015. **43**(2): p. 181-199.
88. Teece, D., M. Peteraf, and S. Leih, *Dynamic capabilities and organizational agility: Risk, uncertainty, and strategy in the innovation economy.* California Management Review, 2016. **58**(4): p. 13-35.
89. Pavlou, P.A. and O.A. El Sawy, *The "third hand": IT-enabled competitive advantage in turbulence through improvisational capabilities.* Information Systems Research, 2010. **21**(3): p. 443-471.
90. Van de Wetering, R., et al., *The Impact of EA-Driven Dynamic Capabilities, Innovativeness, and Structure on Organizational Benefits: A Variance and fsQCA Perspective.* Sustainability, 2021. **13**(10): p. 5414.
91. Cepeda, G. and D. Vera, *Dynamic capabilities and operational capabilities: A knowledge management perspective.* Journal of business research, 2007. **60**(5): p. 426-437.
92. Jayachandran, S., K. Hewett, and P. Kaufman, *Customer response capability in a sense-and-respond era: the role of customer knowledge process.* Journal of the Academy of Marketing Science, 2004. **32**(3): p. 219-233.
93. Grant, R.M., *Prospering in dynamically-competitive environments: Organizational capability as knowledge integration.* Organization science, 1996. **7**(4): p. 375-387.
94. Pucihar, A., *The digital transformation journey: content analysis of Electronic Markets articles and Bled eConference proceedings from 2012 to 2019.* Electronic Markets, 2020: p. 1-9.
95. Im, G. and A. Rai, *Knowledge sharing ambidexterity in long-term interorganizational relationships.* Management science, 2008. **54**(7): p. 1281-1296.
96. Nickerson, J.A. and T.R. Zenger, *A knowledge-based theory of the firm—The problem-solving perspective.* Organization science, 2004. **15**(6): p. 617-632.
97. Tanriverdi, H., *Information technology relatedness, knowledge management capability, and performance of multibusiness firms.* MIS quarterly, 2005: p. 311-334.
98. Kohli, A.K. and B.J. Jaworski, *Market orientation: the construct, research propositions, and managerial implications.* Journal of marketing, 1990. **54**(2): p. 1-18.
99. Turner, A., et al., *A first class knowledge service: developing the National electronic Library for Health.* Health Information & Libraries Journal, 2002. **19**(3): p. 133-145.
100. Stefanelli, M., *Knowledge and process management in health care organizations.* Methods of information in medicine, 2004. **43**(05): p. 525-535.
101. Wang, Y. and T.A. Byrd, *Business analytics-enabled decision-making effectiveness through knowledge absorptive capacity in health care.* Journal of Knowledge Management, 2017.
102. Zahra, S.A. and G. George, *Absorptive capacity: A review, reconceptualization, and extension.* Academy of management review, 2002. **27**(2): p. 185-203.
103. Roberts, N., et al., *Absorptive Capacity and Information Systems Research: Review, Synthesis, and Directions for Future Research.* MIS quarterly, 2012. **36**(2): p. 625-648.
104. Cohen, W.M. and D.A. Levinthal, *Absorptive capacity: A new perspective on learning and innovation.* Administrative science quarterly, 1990: p. 128-152.
105. Liu, H., et al., *The impact of IT capabilities on firm performance: The mediating roles of absorptive capacity and supply chain agility.* Decision Support Systems, 2013. **54**(3): p. 1452-1462.

106. Sambamurthy, V., A. Bharadwaj, and V. Grover, *Shaping agility through digital options: Reconceptualizing the role of information technology in contemporary firms.* MIS quarterly, 2003. **27**(2): p. 237-263.
107. Evans, J.M., A. Brown, and G.R. Baker, *Organizational knowledge and capabilities in healthcare: Deconstructing and integrating diverse perspectives.* SAGE open medicine, 2017. **5**: p. 2050312117712655.
108. Singh, R., et al., *Dynamic capabilities in home health: IT-enabled transformation of post-acute care.* Journal of the Association for Information Systems, 2011. **12**(2): p. 2.
109. Pablo, A.L., et al., *Identifying, enabling and managing dynamic capabilities in the public sector.* Journal of management studies, 2007. **44**(5): p. 687-708.
110. Kislov, R., et al., *Rethinking capacity building for knowledge mobilisation: developing multilevel capabilities in healthcare organisations.* Implementation Science, 2014. **9**(1): p. 166.
111. Gopalakrishna-Remani, V., R.P. Jones, and K.M. Camp, *Levels of EMR adoption in US hospitals: An empirical examination of absorptive capacity, institutional pressures, top management beliefs, and participation.* Information Systems Frontiers, 2019. **21**(6): p. 1325-1344.
112. Vickery, S., et al., *Supply chain information technologies and organisational initiatives: complementary versus independent effects on agility and firm performance.* International Journal of Production Research, 2010. **48**(23): p. 7025-7042.
113. Park, Y., O.A. El Sawy, and P.C. Fiss, *The Role of Business Intelligence and Communication Technologies in Organizational Agility: A Configurational Approach.* Journal of the Association for Information Systems, 2017. **18**(9): p. 648-686.
114. Galbraith, J.R., *Organization design: An information processing view.* Interfaces, 1974. **4**(3): p. 28-36.
115. Roberts, N. and V. Grover, *Investigating firm's customer agility and firm performance: The importance of aligning sense and respond capabilities.* Journal of Business Research, 2012. **65**(5): p. 579-585.
116. Van de Wetering, R. *Achieving digital-driven patient agility in the era of big data*. In the proceedings of *The 20th IFIP Conference e-Business, e-Services, and e-Society I3E2021*. 2021. Galway, Ireland: Springer.
117. El Sawy, O.A. and P.A. Pavlou, *IT-enabled business capabilities for turbulent environments.* MIS Quarterly Executive (2008), 2008. **7**(3): p. 139-150.
118. Mithas, S., et al., *Information technology and firm profitability: mechanisms and empirical evidence.* 2012.
119. Setia, P., V. Venkatesh, and S. Joglekar, *Leveraging digital technologies: How information quality leads to localized capabilities and customer service performance.* Mis Quarterly, 2013. **37**(2).
120. Tippins, M.J. and R.S. Sohi, *IT competency and firm performance: is organizational learning a missing link?* Strategic management journal, 2003. **24**(8): p. 745-761.
121. Ravichandran, T., *Exploring the relationships between IT competence, innovation capacity and organizational agility.* The Journal of Strategic Information Systems, 2018. **27**(1): p. 22-42.
122. Chaudhry, B., et al., *Systematic review: impact of health information technology on quality, efficiency, and costs of medical care.* Annals of Internal Medicine, 2006. **144**(10): p. 742–752.
123. Bhattacherjee, A. and N. Hikmet, *Physicians' resistance toward healthcare information technology: a theoretical model and empirical test.* European Journal of Information Systems, 2007. **16**(6): p. 725-737.
124. Gregory, R.W., et al., *Paradoxes and the nature of ambidexterity in IT transformation programs.* Information Systems Research, 2015. **26**(1): p. 57-80.
125. Jaana, M., M.M. Ward, and J.A. Bahensky, *EMRs and clinical IS implementation in hospitals: a statewide survey.* The Journal of Rural Health, 2012. **28**(1): p. 34-43.
126. Chen, Y., H. Liu, and M. Chen, *Achieving Novelty and Efficiency in Business Model Design: Striking a Balance between IT Exploration and Exploitation.* Information & Management, 2020: p. 103268.
127. Voudouris, I., et al., *Effectiveness of technology investment: Impact of internal technological capability, networking and investment's strategic importance.* Technovation, 2012. **32**(6): p. 400-414.
128. Mao, H., S. Liu, and J. Zhang, *How the effects of IT and knowledge capability on organizational agility are contingent on environmental uncertainty and information intensity.* Information Development, 2015. **31**(4): p. 358-382.
129. Benitez, J., et al., *IT-enabled knowledge ambidexterity and innovation performance in small US firms: The moderator role of social media capability.* Information & Management, 2018. **55**(1): p. 131-143.
130. Nazir, S. and A. Pinsonneault, *IT and firm agility: an electronic integration perspective.* Journal of the Association for Information Systems, 2012. **13**(3): p. 2.
131. Acur, N., et al., *Exploring the impact of technological competence development on speed and NPD program performance.* Journal of product innovation management, 2010. **27**(6): p. 915-929.

132. Zhou, K.Z. and F. Wu, *Technological capability, strategic flexibility, and product innovation.* Strategic Management Journal, 2010. **31**(5): p. 547-561.
133. Wang, Y., H.-P. Lo, and Y. Yang, *The constituents of core competencies and firm performance: evidence from high-technology firms in China.* Journal of Engineering and Technology Management, 2004. **21**(4): p. 249-280.
134. Coombs, J.E. and P.E. Bierly III, *Measuring technological capability and performance.* R&D Management, 2006. **36**(4): p. 421-438.
135. Ritter, T. and C.L. Pedersen, *Digitization capability and the digitalization of business models in business-to-business firms: Past, present, and future.* Industrial Marketing Management, 2019.
136. Wade, M. and J. Hulland, *Review: The resource-based view and information systems research: Review, extension, and suggestions for future research.* MIS quarterly, 2004. **28**(1): p. 107-142.
137. Van de Wetering, R., J. Versendaal, and P. Walraven. *Examining the relationship between a hospital's IT infrastructure capability and digital capabilities: a resource-based perspective*. In the proceedings of the Twenty-fourth Americas Conference on Information Systems (AMCIS). 2018. New Orleans: AIS.
138. Tzokas, N., et al., *Absorptive capacity and performance: The role of customer relationship and technological capabilities in high-tech SMEs.* Industrial Marketing Management, 2015. **47**: p. 134-142.
139. Wang, Y., L. Kung, and T.A. Byrd, *Big data analytics: Understanding its capabilities and potential benefits for healthcare organizations.* Technological Forecasting and Social Change, 2018. **126**: p. 3-13.
140. Wamba, S.F., et al., *Big data analytics and firm performance: Effects of dynamic capabilities.* Journal of Business Research, 2017. **70**: p. 356-365.
141. Delaney, J.T. and M.A. Huselid, *The impact of human resource management practices on perceptions of organizational performance.* Academy of Management journal, 1996. **39**(4): p. 949-969.
142. Hobbs, K.W., et al., *Incorporating Information From Electronic and Social Media Into Psychiatric and Psychotherapeutic Patient Care: Survey Among Clinicians.* Journal of medical Internet research, 2019. **21**(7): p. e13218.
143. Pronovost, P.J., et al., *Evaluation of the culture of safety: survey of clinicians and managers in an academic medical center.* BMJ Quality & Safety, 2003. **12**(6): p. 405-410.
144. DesRoches, C.M., et al., *Electronic health records in ambulatory care—a national survey of physicians.* New England Journal of Medicine, 2008. **359**(1): p. 50-60.
145. Zahra, S.A. and J.G. Covin, *Business strategy, technology policy and firm performance.* Strategic management journal, 1993. **14**(6): p. 451-478.
146. Berg, N., *Non-response bias.* 2005.
147. Hikmet, N. and S.K. Chen, *An investigation into low mail survey response rates of information technology users in health care organizations.* International journal of medical informatics, 2003. **72**(1-3): p. 29-34.
148. Richardson, H.A., M.J. Simmering, and M.C. Sturman, *A tale of three perspectives: Examining post hoc statistical techniques for detection and correction of common method variance.* Organizational Research Methods, 2009. **12**(4): p. 762-800.
149. Podsakoff, P.M., et al., *Common method biases in behavioral research: A critical review of the literature and recommended remedies.* Journal of applied psychology, 2003. **88**(5): p. 879.
150. Ringle, C.M., S. Wende, and J.-M. Becker, *SmartPLS 3.* Boenningstedt: SmartPLS GmbH, http://www. smartpls. com, 2015.
151. Rigdon, E.E., M. Sarstedt, and C.M. Ringle, *On comparing results from CB-SEM and PLS-SEM: five perspectives and five recommendations.* Marketing Zfp, 2017. **39**(3): p. 4-16.
152. Hair Jr, J.F., et al., *A primer on partial least squares structural equation modeling (PLS-SEM)*. 2016: Sage Publications.
153. Hair Jr, J.F., et al., *Advanced issues in partial least squares structural equation modeling*. 2017: SAGE Publications.
154. Lowry, P.B. and J. Gaskin, *Partial least squares (PLS) structural equation modeling (SEM) for building and testing behavioral causal theory: When to choose it and how to use it.* IEEE transactions on professional communication, 2014. **57**(2): p. 123-146.
155. Chin, W., *Issues and opinion on structural equation modeling.* Management Information Systems Quarterly, 1998. **22**(1): p. 7–16.
156. Hair, J.F., C.M. Ringle, and M. Sarstedt, *PLS-SEM: Indeed a silver bullet.* Journal of Marketing theory and Practice, 2011. **19**(2): p. 139-152.
157. Faul, F., et al., *Statistical power analyses using G* Power 3.1: Tests for correlation and regression analyses.* Behavior research methods, 2009. **41**(4): p. 1149-1160.
158. Fornell, C. and D. Larcker, *Evaluating structural equation models with unobservable variables and measurement error.* Journal of Marketing Research, 1981. **18**(1): p. 39–50.

159. Farrell, A.M., *Insufficient discriminant validity: A comment on Bove, Pervan, Beatty, and Shiu (2009)*. Journal of Business Research, 2010. **63**(3): p. 324-327.
160. Henseler, J., C.M. Ringle, and M. Sarstedt, *A new criterion for assessing discriminant validity in variance-based structural equation modeling*. Journal of the Academy of Marketing Science, 2015. **43**(1): p. 115-135.
161. Kock, N. and G. Lynn, *Lateral collinearity and misleading results in variance-based SEM: An illustration and recommendations*. 2012.
162. Hu, L.t. and P.M. Bentler, *Cutoff criteria for fit indexes in covariance structure analysis: Conventional criteria versus new alternatives*. Structural equation modeling: a multidisciplinary journal, 1999. **6**(1): p. 1-55.
163. Chin, W., *The partial least squares approach to structural equation modeling*, in *Modern Methods for Business Research*, G.A. Marcoulides, Editor. 1998, Lawrence Erlbaum Associates: Mahwah, N.J. p. 295–336.
164. Hayes, A.F., *Introduction to mediation, moderation, and conditional process analysis: A regression-based approach*. 2013: Guilford Press.
165. Agarwal, R., et al., *Research commentary—The digital transformation of healthcare: Current status and the road ahead*. Information Systems Research, 2010. **21**(4): p. 796-809.
166. Herrmann, M., et al., *Digital transformation and disruption of the health care sector: internet-based observational study*. Journal of medical internet research, 2018. **20**(3): p. e104.
167. McGrail, K.M., M.A. Ahuja, and C.A. Leaver, *Virtual visits and patient-centered care: results of a patient survey and observational study*. Journal of medical Internet research, 2017. **19**(5): p. e177.
168. Pang, M.-S., G. Lee, and W.H. DeLone, *IT resources, organizational capabilities, and value creation in public-sector organizations: a public-value management perspective*. Journal of Information Technology, 2014. **29**(3): p. 187-205.
169. Gerybadze, A., *Technological competence assessment within the firm: applications of competence theory to managerial practice*. 1998: Forschungsstelle Internat. Management und Innovation.
170. Ashrafi, N., et al. *Boosting enterprise agility via IT knowledge management capabilities*. in *Proceedings of the 39th Annual Hawaii International Conference on System Sciences (HICSS'06)*. 2006. IEEE.
171. Schryen, G., *Revisiting IS business value research: what we already know, what we still need to know, and how we can get there*. European Journal of Information Systems, 2013. **22**(2): p. 139-169.
172. Sabherwal, R. and A. Jeyaraj, *Information Technology Impacts on Firm Performance: An Extension of Kohli and Devaraj (2003)*. MIS quarterly, 2015. **39**(4): p. 809-836.
173. Gray, C.S., *Seeking Meaningful Innovation: Lessons Learned Developing, Evaluating, and Implementing the Electronic Patient-Reported Outcome Tool*. Journal of Medical Internet Research, 2020. **22**(7): p. e17987.
174. Van Velthoven, M.H. and C. Cordon, *Sustainable adoption of digital health innovations: perspectives from a stakeholder workshop*. Journal of medical Internet research, 2019. **21**(3): p. e11922.
175. Papoutsi, C., et al., *Putting the social back into sociotechnical: Case studies of co-design in digital health*. Journal of the American Medical Informatics Association, 2020.
176. Keesara, S., A. Jonas, and K. Schulman, *Covid-19 and health care's digital revolution*. New England Journal of Medicine, 2020. **382**(23): p. e82.
177. Brosius, M., et al. *Enterprise Architecture Assimilation: An Institutional Perspective*. 2018. Association for Information Systems.
178. Hazen, B.T., et al., *Enterprise architecture: A competence-based approach to achieving agility and firm performance*. Management, 2017. **193**: p. 566-577.
179. Shanks, G., et al., *Achieving benefits with enterprise architecture*. The Journal of Strategic Information Systems, 2018. **27**(2): p. 139-156.
180. Schmidt, C. and P. Buxmann, *Outcomes and success factors of enterprise IT architecture management: empirical insight from the international financial services industry*. European Journal of Information Systems, 2011. **20**(2): p. 168-185.
181. Wu, S.P.-J., D.W. Straub, and T.-P. Liang, *How information technology governance mechanisms and strategic alignment influence organizational performance: Insights from a matched survey of business and IT managers*. Mis Quarterly, 2015. **39**(2): p. 497-518.
182. Egener, B.E., et al., *The charter on professionalism for health care organizations*. Academic Medicine, 2017. **92**(8): p. 1091.
183. Donelan, K., et al., *Perspectives of physicians and nurse practitioners on primary care practice*. New England Journal of Medicine, 2013. **368**(20): p. 1898-1906.